\documentclass[twocolumn,showpacs,preprintnumbers,amsmath,amssymb]{revtex4}

\usepackage{graphicx}% Include figure files
\usepackage{dcolumn}% Align table columns on decimal point
\usepackage{bm}% bold math
\usepackage{dsfont}
\begin{document}

\title{Excited hadrons on the lattice: Mesons}
\author{Tommy Burch,$^1$ 
Christof Gattringer,$^2$ 
Leonid Ya.\ Glozman,$^2$ 
Christian Hagen,$^1$ 
C.\ B.\ Lang,$^2$ and 
Andreas Sch\"afer$^1$\\
(BGR [Bern-Graz-Regensburg] Collaboration)}
\vskip1mm
\affiliation{$^1$Institut f\"ur Theoretische Physik, Universit\"at
Regensburg, D-93040 Regensburg, Germany}
\affiliation{$^2$Institut f\"ur Physik, FB Theoretische Physik, Universit\"at
Graz, A-8010 Graz, Austria}

\begin{abstract}
We present results for masses of excited mesons from quenched calculations 
using chirally improved quarks at pion masses down to 350 MeV. 
The key features of our analysis are the use of a matrix of correlators 
from various source and sink operators and a basis which includes quark 
sources with different spatial widths, thereby improving overlap with states 
exhibiting radial excitations.
\end{abstract}
\maketitle

\section{Introduction}\label{SectIntroduction}

Ground-state hadron spectroscopy has been at the forefront of the studies 
performed by the lattice QCD community. 
The exponential decay behavior of the mass eigenstate contributions to 
Euclidean-time correlators makes extraction of the lightest state 
straightforward: at large time separations, the ground-state mass naturally 
dominates the correlator, which then displays a single exponential in time. 
Excited states, however, appear as subleading exponentials, which in practice 
are rather difficult to separate, not only from the ground-state contribution, 
but also from those due to other excited states.

A few methods are currently in use to deal with this problem. 
A straightforward fit to a finite sum of exponentials is sometimes possible, 
usually when high statistics are available. 
Constrained fitting \cite{constrainedfits} 
and the Maximum Entropy Method \cite{maximumentropy} 
are also options, but these, like the first method, can encounter problems 
when faced with states which lie close in mass, or when unphysical 
contributions appear due to quenching (ghosts).

We employ a different method, one based upon the variational method devised 
by Michael \cite{Mi85}, and later elaborated by L\"uscher and Wolff 
\cite{LuWo90}. 
The distinguishing characteristic of our approach is the use of several 
hadron interpolators which contain quark wavefunctions covariantly smeared to 
approximate Gaussians of different widths \cite{BuGaGl04}. 
Using such a basis of source and sink operators, it is our plan to improve 
overlap with physical states which may involve a radial excitation. 
In a recent paper \cite{BuGaGl05}, it has been demonstrated that this 
method also clearly separates ghost contributions.

In the present work, we describe the method in detail and report results for 
the meson sector. 
Another paper will contain our latest results for baryons.

\section{The method}\label{SectMethod}
Our calculation is based upon the variational method \cite{Mi85,LuWo90}. 
The central idea is to use several different interpolators 
$O_i, i = 1, \ldots \, N$ 
with the quantum numbers of the desired state and to compute all cross 
correlations 
\begin{equation}
C(t)_{ij} \; = \; \langle \, O_i(t) \, \overline{O}_j(0) \, \rangle \; . 
\label{corrmatdef}
\end{equation}
In Hilbert space these correlators have the decomposition 
\begin{equation}
C(t)_{ij} \; = \; \sum_n \langle \, 0 \, | \, O_i \, | \, n \, \rangle 
\langle \, n \, | \, O_j^\dagger \, | \, 0 \, \rangle \, e^{-t \, M_n}  \; . 
\label{corrmatrix}
\end{equation}
Using the factorization of the amplitudes one can show \cite{LuWo90} 
that the eigenvalues 
$\lambda^{(k)}(t)$ of the generalized eigenvalue problem 
\begin{equation}
C(t) \, \vec{v}^{(k)} \; \; = \; \; \lambda^{(k)}(t) \, C(t_0) \, 
\vec{v}^{(k)} \; , 
\label{generalized}
\end{equation}
behave as 
\begin{equation}
\lambda^{(k)}(t) \; \propto \; e^{-t \, M_k} \,[ \, 1 +
{\cal O}(e^{-t \, \Delta M_k}) \,] \; , 
\label{eigenvaluedecay}
\end{equation}
where $M_k$ is the mass of the $k$-th state and $\Delta M_k$ is the difference 
to the mass closest to $M_k$ \cite{WeMo06}. In Eq.\ (\ref{generalized}) 
the eigenvalue problem is normalized with respect to a timeslice $t_0 < t$.

Equation (\ref{eigenvaluedecay}) shows that the eigenvalues each decay 
with their own mass: The largest eigenvalue decays with the mass of the 
ground state, the second largest eigenvalue with the mass of the first 
excited state, etc. Thus, the variational method allows one to decompose 
the signal into those for ground and excited states, as well as ghost 
contributions \cite{BuGaGl05}, and therefore, simple, stable two-parameter 
fits become possible.

We remark that also for the regular eigenvalue problem, 
\begin{equation}
C(t) \, \vec{v}^{(k)} \; \; = \; \; \lambda^{(k)}(t) \,  
\vec{v}^{(k)} \; , 
\label{regular}
\end{equation}
a behavior of the type (\ref{eigenvaluedecay}) can be shown \cite{LuWo90}. 
We find that in a practical implementation with our 
sources the results and the quality of the data are essentially unchanged 
whether we use the regular or the generalized eigenvalue problem.

The key to a successful application of the variational method is the choice 
of the basis interpolators. They should be linearly independent, and 
as orthogonal as possible, while at the same time, able to represent 
the physical state as well as possible. 
Furthermore, they should be numerically cheap to implement.

For meson spectroscopy it is well known that different Dirac structures can 
be used to construct interpolators with the quantum numbers one is interested 
in. In this paper we consider interpolators of the form 
\begin{equation}
O \; = \; \overline{\psi}^{(f_1)} \, \Gamma \, \psi^{(f_2)} \; , 
\label{basicmeson}
\end{equation}
where $f_i$ are flavor labels. We use matrix/vector 
notation for Dirac and color indices and $\Gamma$ is an element of the Clifford 
algebra. Depending on what combination of gamma-matrices one uses for 
$\Gamma$, the interpolator $O$ will have different quantum numbers. 
We list these in Table~\ref{gammatable}.

\begin{table}[t]
\caption{
Quantum numbers of the interpolators (\protect{\ref{basicmeson}}) for 
different choices of $\Gamma$. We remark that the classification 
with respect to $C$ is for flavor degenerate interpolators only.}
\label{gammatable}
\begin{center}
\begin{tabular}{llll}
\hline \hline
state \hspace{13mm} & $J^{PC}$ \hspace{5mm} & 
$\Gamma$ \hspace{13mm} & particles \hspace{10mm}\\
\hline 
scalar & $0^{++}$ & $\mathds{1}$ & $a_0$ \\
pseudoscalar & $0^{-+}$ & $\gamma_5 \, , \, \gamma_4 \gamma_5$ & $\pi, K$ \\
vector & $1^{--}$ & $\gamma_i \, , \, \gamma_4 \gamma_i$ & $\rho, K^*, \phi $ \\
pseudovector & $1^{++}$ & $\gamma_i \gamma_5$ & $a_1$ \\
pseudovector & $1^{+-}$ & $\gamma_i \gamma_j$ & $b_1$ \\
\hline \hline
\end{tabular}
\end{center}
\end{table}

However, it is well known (see, e.g., \cite{blumetal,BrCrGa04}) 
that correlating operators with different Dirac structures alone 
does not provide a sufficient basis to obtain good overlap with excited 
states. For many excited hadrons the spatial wavefunctions are 
expected to have nodes. In \cite{BuGaGl04,BuGaGl04b} it was proposed, and 
tested on small lattices, to use Jacobi smeared quark sources 
of different widths to allow for nodes in the radial wavefunction. 
Other lattice efforts \cite{Al93,LHP05} have also seen the need of using 
spatially extended operators.

In a lattice spectroscopy calculation the hadron correlators are built 
from quark propagators $D^{-1}$ acting on a source $s$, 
\begin{equation}
\sum_{\vec{y},\rho,c}
D^{-1}(\vec{x}, t \mid \vec{y}, 0)_{\beta \, \rho \atop b \, c} \; 
s^{(\alpha, a)} (\vec{y},0)_{\rho \atop c} \; . 
\label{diraconsource}
\end{equation}
If the source is point-like, i.e., $s = s_0$, with 
\begin{equation}
s^{(\alpha, a)}_0 (\vec{y},0)_{\rho \atop c} \; \;  = \; \;
\delta(\vec{y}, \vec{0}) \; \delta_{\rho \, \alpha} \; \delta_{c \, a} \; , 
\label{pointsource}
\end{equation}
then the two quarks in (\ref{basicmeson}) both sit on the same lattice 
site. Certainly this is not a very physical assumption.

The idea of Jacobi smearing \cite{jacobi1,jacobi2} is to 
create an extended source by iteratively applying the hopping part of 
the Wilson term within the timeslice of the source: 
\begin{eqnarray}
s^{(\alpha,a)} &  =  & M \; s_0^{(\alpha,a)} 
\; \; \; , \; \; \; \; 
M \, = \, \sum_{n=0}^N \, \kappa^n \, H^n \; , 
\nonumber
\\
H(\vec{x},\vec{y}\,) &  =  & \sum_{j = 1}^3
\Big[ \, U_j(\vec{x},0) \, \delta(\vec{x} + \hat{j\,}, \vec{y}\,)
\nonumber
\\
&& \qquad+ \; \,
 U_j(\vec{x} \! - \! \hat{j\,},0)^\dagger \, 
\delta(\vec{x} - \hat{j\,}, \vec{y}\,) 
\, \Big] \; . 
\label{jacobismear}
\end{eqnarray}
Applying the inverse Dirac operator as shown in (\ref{diraconsource}) 
connects the source at timeslice $t = 0$ to the lattice points at 
timeslice $t$. There an extended sink may be created by again applying 
the smearing operator $M$.

\begin{figure}[t]
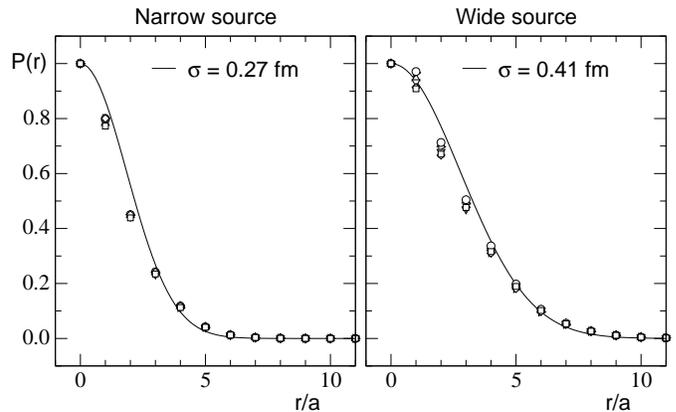

\vspace*{1mm}
\hspace*{-3mm}
\includegraphics*[height=5.4cm]{source_narrow.eps} 
\hspace{-2mm}
\includegraphics*[height=5.4cm]{source_wide.eps}
\caption{Profiles $P(r)$ of the narrow and wide source 
as a function of the radius $r$ (for the $\beta = 7.90$ lattice). 
The symbols are our data points. The curves 
are the target Gaussian distributions (with half width $\sigma$) 
which we approximate by the profiles $P(r)$.
\label{profilefig}}
\end{figure}

Jacobi smearing has two free parameters, the hopping parameter 
$\kappa$ and the number of smearing steps $N$. They can be used 
to create sources and sinks with approximately Gaussian profiles 
of different widths. 
In Fig.\ \ref{profilefig} we show two such profiles $P(r)$ as a function 
of the radius $r$. For mapping these profiles we use the definition 
\begin{equation}
P(r) \; = \; \sum_{\vec{y}} \delta \Big( \, |\vec{y} \,| -  r\, \Big) \, 
\sum_b \, 
\Big| s^{(\alpha, a)} (\vec{y},0)_{\alpha \atop b} \Big|\; . 
\label{profile}
\end{equation}

\begin{table}[b]
\caption{
Parameters of our simulation. We list the lattice size, the inverse coupling 
$\beta$, the number of configurations, the lattice spacing $a$, the 
cutoff $a^{-1}$, and the smearing parameters $N$ and $\kappa$ for 
the narrow and wide sources.}
\label{parametertable}
\begin{center}
\begin{tabular}{ccccccc}
\hline \hline
size & $\beta$ & confs. & $a$[fm] & $a^{\!-1}$[MeV] & $N \,(n,w)$ 
& $\kappa \,(n,w)$\\
\hline 
$20^3\! \times \! 32$ & 8.15 & 100 & 0.119 & 1680 & 22, 62 & 0.21, 0.1865 \\
$16^3\! \times \! 32$ & 7.90 & 100 & 0.148 & 1350 & 18, 41 & 0.21, 0.1910 \\
\hline \hline
\end{tabular}
\end{center}
\end{table}

\begin{figure*}[t]
\vspace*{0mm}
\hspace*{0mm}
\includegraphics*[width=14.5cm,clip,angle=270]{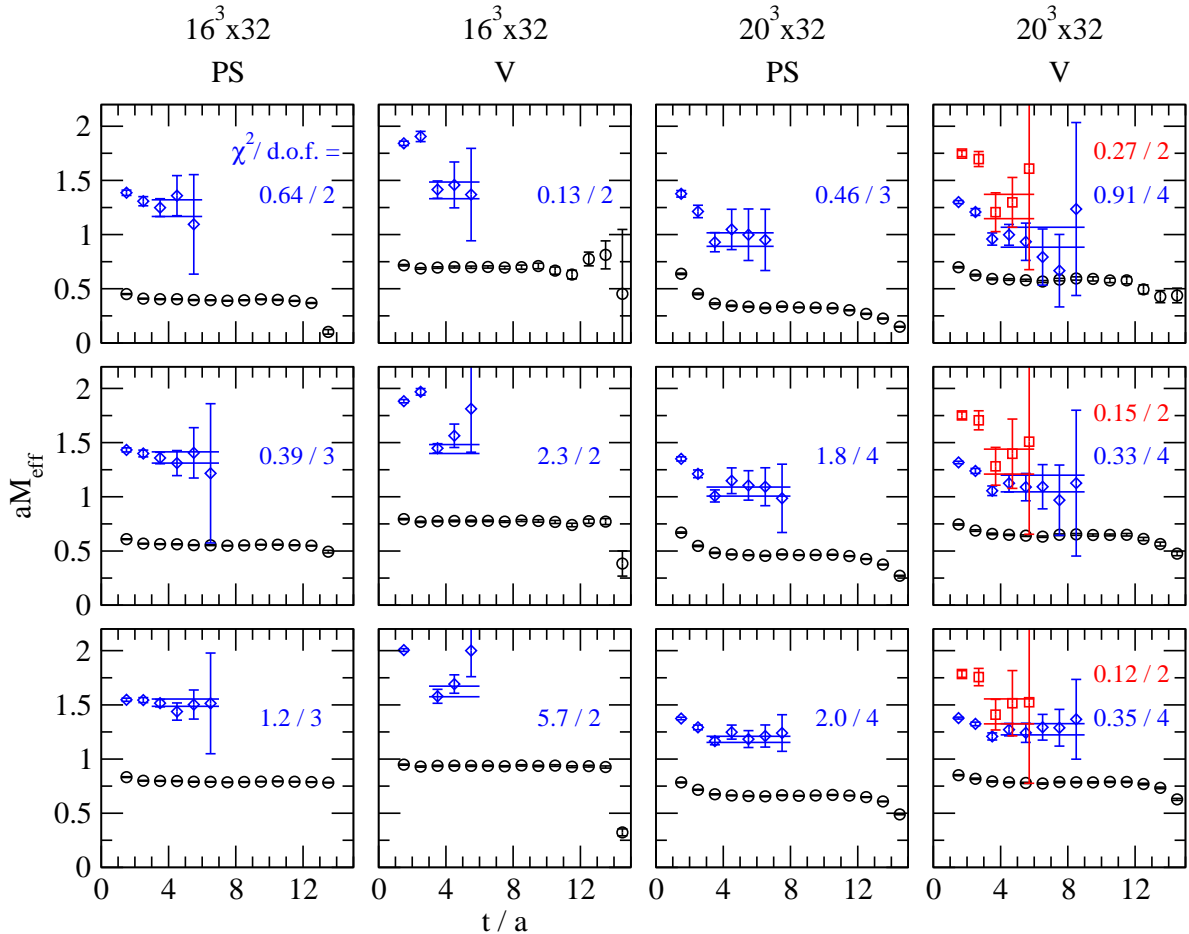}
\caption{ Effective mass plots for pseudoscalar (PS) and vector (V) mesons 
from our coarse ($16^3 \times 32$, $a=0.148$ fm, with $am_q=0.05,0.1,0.2$ 
from top to bottom) and fine ($20^3 \times 32$, $a=0.119$ fm, with 
$am_q=0.04,0.08,0.16$) lattices. 
Both ground and excited states are shown, along with the $M\pm\sigma_M^{}$ 
results (horizontal lines) from correlated fits to the corresponding time 
intervals. 
The numbers in the plots are the $\chi^2$ per degree of freedom values for 
the fits to the excited-state plateaus.
\label{eff_mass_ps_vt}}
\end{figure*}

The parameters $\kappa$ and $N$ were adjusted such that the narrow 
source approximates a Gaussian with a half width (i.e., standard deviation) 
of 0.27 fm and the wide source a Gaussian with a half width of 0.41 fm. 
The corresponding smearing parameters are listed in 
Table \ref{parametertable}. 
The values of the smearing parameters are chosen such that simple linear 
combinations of our two sources allows one to approximate the ground- and 
first-excited state wavefunctions of the 3-d harmonic oscillator 
(with a half width of $\sigma = 0.33$ fm for the ground state). 
%These linear combinations are 
%($n$ = narrow, $w$ = wide): 
%\begin{eqnarray}
%\mbox{ground state} & \sim & 0.6 \, n \; + \; 0.4 \, w \; ,
%\nonumber \\
%\mbox{first excited} & \sim & 2.2 \, n \; - \; 1.2 \, w \; . 
%\label{sourcelincombs}
%\end{eqnarray}
We stress that we do not impose such linear combinations, 
but rather use the simple $n$, $w$ sources for 
the basis interpolators in the variational method and leave it to the 
simulation to determine the physical superpositions.
%The comparison to the harmonic oscillator wavefunctions is only useful for 
%choosing reasonable values for the smearing parameters $N, \, \kappa$.

The interpolators which we actually consider in our correlation matrix 
are flavor triplet operators of the form, 
\begin{equation}
O^\Gamma \; = \; \overline{u} \, \Gamma \, d \; . 
\label{tripcorr}
\end{equation}
This avoids the need to calculate disconnected pieces. 
For a given Clifford element $\Gamma$ both $\overline{u}$ and $d$ sources 
can be wide or narrow which gives the 4 possibilities (the subscripts 
$n, w$ denote the type of smearing used) 
\begin{eqnarray}
O^\Gamma_{nn} \, = \, \overline{u}_n \, \Gamma \, d_n 
& \; , \; &
O^\Gamma_{wn} \, = \, \overline{u}_w \, \Gamma \, d_n \; , 
\nonumber \\
O^\Gamma_{nw} \, = \, \overline{u}_n \, \Gamma \, d_w 
& \; , \; &
O^\Gamma_{ww} \, = \, \overline{u}_w \, \Gamma \, d_w \; . 
\end{eqnarray}
In addition, we have two choices for $\Gamma$ in the vector and 
pseudoscalar sectors such that for these cases we have a basis of 8 
interpolators. 
For scalars and pseudovectors we restrict ourselves to 4 interpolators.

In the case of degenerate quark masses, $O^\Gamma_{wn}$ and 
$O^\Gamma_{nw}$ give rise to identical correlators. This fact 
reduces the correlation matrices to $6 \times 6$ for vector and pseudoscalar states and to 
$3 \times 3$ for scalar and pseudovector states. 
For the strange mesons, obtained by replacing the down quark in 
(\ref{tripcorr}) by a strange quark, the degeneracy is lifted and we are free 
to work with the full $8 \times 8$ and $4 \times 4$ correlation matrices.

Our quenched study is done with the L\"uscher-Weisz gauge action 
\cite{Luweact} at two different values of the inverse gauge coupling 
$\beta$. The lattice spacing $a$ was determined in \cite{scale} 
using the Sommer parameter. We use lattices of two different sizes, 
such that the spatial extent in physical units is kept constant 
at 2.4 fm. This allows us to compare the results at two different 
values of the cutoff, $a^{-1} = 1680$ MeV and $a^{-1} = 1350$ MeV. 
For the fermions we use the chirally improved Dirac operator 
\cite{chirimp}. It is an approximation of a solution of the Ginsparg-Wilson 
equation \cite{GiWi82}, with good chiral behavior \cite{bgrlarge}. 
The parameters of our calculation are collected in Table~\ref{parametertable}. 
We determine the strange quark mass via interpolations in the heavy-quark 
mass which match the (light-quark mass extrapolated) pseudoscalar K meson mass 
to the physical value. 
(These configurations have also been used in another study \cite{GaHuLa05} 
to determine low energy constants.)

We remark that the CI operator has one term which is next-to-nearest 
neighbor. 
This has to be kept in mind when selecting fit ranges for masses. 
Exactly what is to be considered a safe minimum time separation before 
fitting is not, however, a simple matter. 
In the present case, we limit ourselves to values larger than $\Delta t=2$.

\section{Results}

\subsection{Effective masses and fit ranges}

Since this is a report on lattice spectroscopy, we begin the discussion of 
our results in the standard manner: with effective masses. 
Having performed the necessary diagonalizations of the correlator matrices, 
we determine the effective masses from ratios of eigenvalues on 
adjacent timeslices: 
\begin{equation}
aM_{eff}^{(k)} \, \Big(t + \frac{1}{2}\Big) \; = \; 
\ln \left( \frac{\lambda^{(k)}(t)}
{\lambda^{(k)}(t+1)} \right) \; .
\end{equation}
Errors for such quantities are determined via a single-elimination jackknife 
procedure.

We note that the quality of the plateaus which we obtain here can depend 
upon the basis chosen. This is simply due to the fact that the various 
interpolators have different overlap with the states which contribute to 
the higher-order corrections in Eq.\ (\ref{eigenvaluedecay}) 
(e.g., see Ref.\ \cite{Al93}). 
An appropriate choice of operator combination (including the possibility of 
further limiting the basis) can therefore minimize such corrections, 
improving the plateaus in effective mass (and correspondingly, in the 
eigenvector components).

In Fig.\ \ref{eff_mass_ps_vt}, we present our effective mass plots for what 
we consider to be our optimal operator combinations 
($n \Gamma n$, $n \gamma_4 \Gamma n$, $n \gamma_4 \Gamma w$, $w \gamma_4 \Gamma w$ 
for pseudoscalar and vector mesons on our coarse lattices; 
$n \gamma_4 \Gamma n$, $n \gamma_4 \Gamma w$, $w \gamma_4 \Gamma w$ 
for the same mesons on our fine lattices; and 
$n \Gamma n$, $n \Gamma w$, $w \Gamma w$ otherwise). 
Shown are results for the pseudoscalar and vector mesons from three quark 
masses on both sets of lattices. 
In each case, the effective masses from the largest two eigenvalues 
(largest three for the vectors on the fine lattice) are displayed. 
The horizontal lines in the plots mark the $M\pm\sigma_M^{}$ values which 
arise from correlated fits to the corresponding time intervals 
(we fit only when we see a plateau of at least three successive 
effective-mass points). 
Also given in the plots are the $\chi^2/d.o.f.$ values for the fits to the 
excited-state plateaus. 
Here, one can see questionable plateaus, and hence poor fits, for the 
coarse-lattice excited vectors at high quark mass, but this is the only place 
where such high $\chi^2/d.o.f.$ values occur. 
For all other fits, $\chi^2/d.o.f. \lesssim 1$.

\begin{figure}[t]
\vspace*{0mm}
\hspace*{0mm}
\includegraphics*[width=7.5cm]{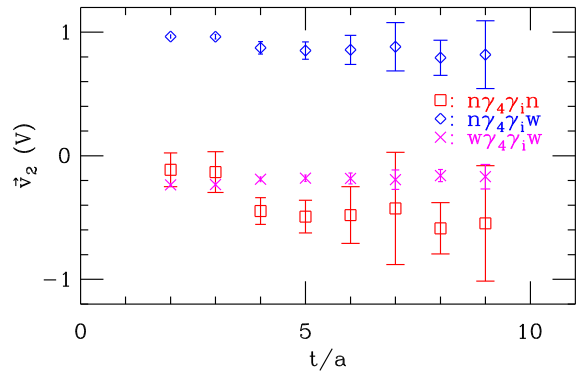}
\caption{ Eigenvector for the first-excited vector meson on our $20^3 \times 32$ 
fine lattices ($a \approx 0.12$ fm, $am_q = 0.06$, $t_0=1a$). The eigenvector on 
each time slice is normalized such that $|\vec v|=1$. 
There is a clear jump in the relative components from $t=3a$ to $t=4a$, 
a plateau forming afterwards. 
In this situation, we fit starting at $t=4a$, even though the corresponding 
effective mass plot shows a plateau starting sooner. We apply similar 
restrictions to all our fit ranges.
\label{ev2_v_20}}
\end{figure}

To be sure that we are dealing (within statistical errors) with a single 
mass eigenstate, we check that the corresponding eigenvectors remain constant 
over the same region where we fit the eigenvalues for the mass. 
An example of such a test is shown in Fig.\ \ref{ev2_v_20}, 
where we display the second eigenvector for one of the vector mesons on our 
fine lattices. 
It is obvious here that plateaus for this state cannot be trusted before 
$t=4a$, even though when looking at the corresponding effective masses, one 
is tempted to start the fit at $t=3a$ (see the right-most column in 
Fig.\ \ref{eff_mass_ps_vt}). We make sure that all our fit 
ranges obey (within errors) such restrictions. 
(This procedure accounts for the enlargement of error bars for the 
aforementioned state since we reported preliminary results for the mesons 
\cite{BuHaHi05}. We are simply being more conservative now.)

\begin{figure}[t]
\vspace*{0mm}
\hspace*{0mm}
\includegraphics*[width=7.5cm]{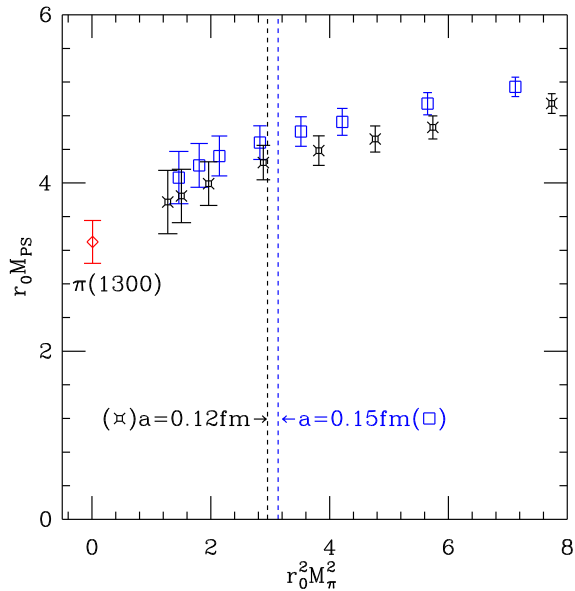}
\caption{ Excited-state pseudoscalar (PS) masses vs.\ $M_\pi^2$ for both 
lattice spacings. The quark masses are degenerate. 
All quantities are in units of the Sommer parameter, $r_0$. 
The diamond represents the experimental point and the vertical lines mark 
the values of $r_0^2 M_\pi^2$ corresponding to the physical strange quark 
mass.
\label{ps_ud_vs_ps2}}
\end{figure}

\begin{figure*}[t]
\vspace*{0mm}
\hspace*{0mm}
\includegraphics*[width=7.5cm]{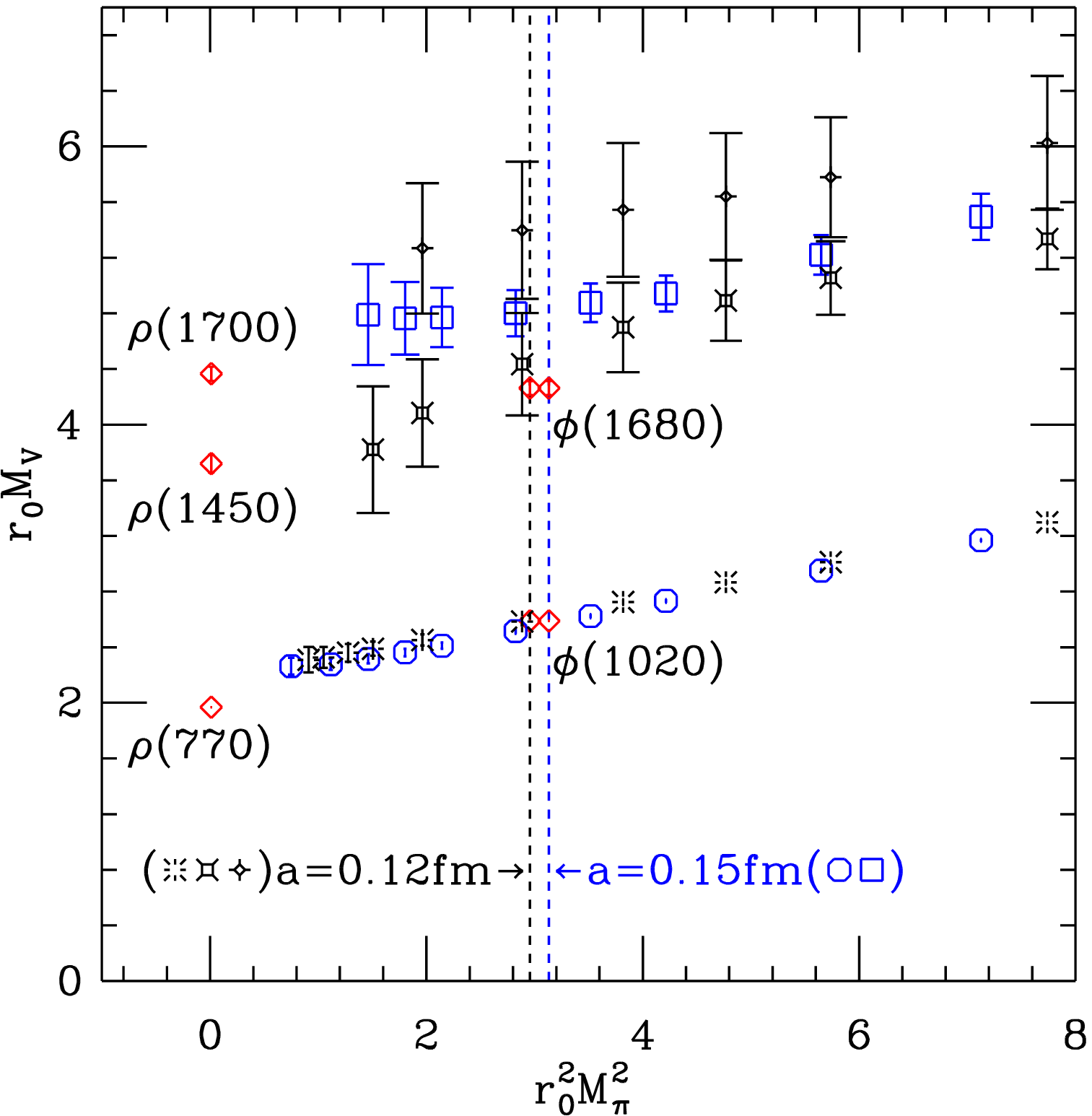}
\hspace*{5mm}
\includegraphics*[width=7.5cm]{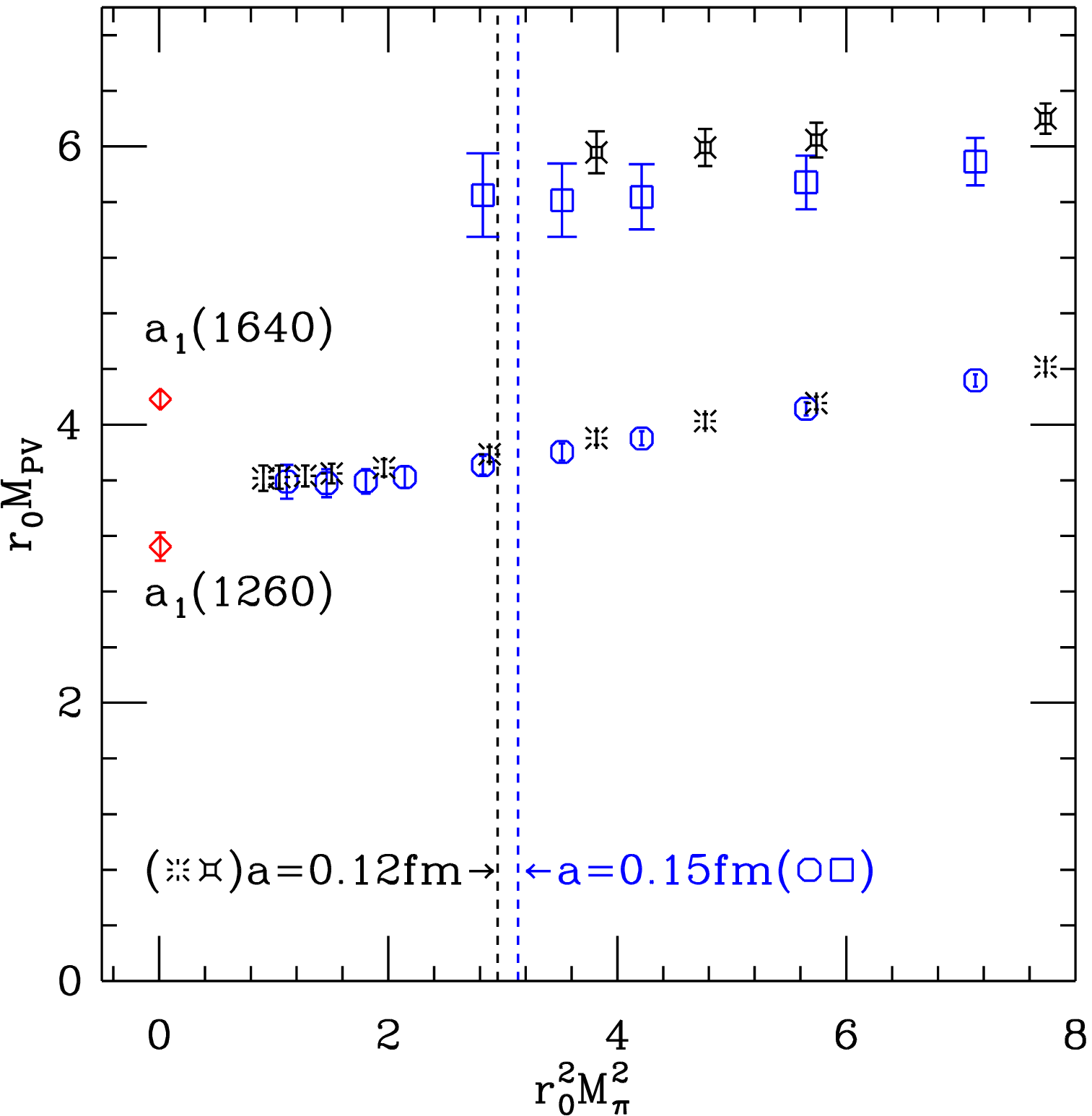}\\
\vspace*{5mm}
\includegraphics*[width=7.5cm]{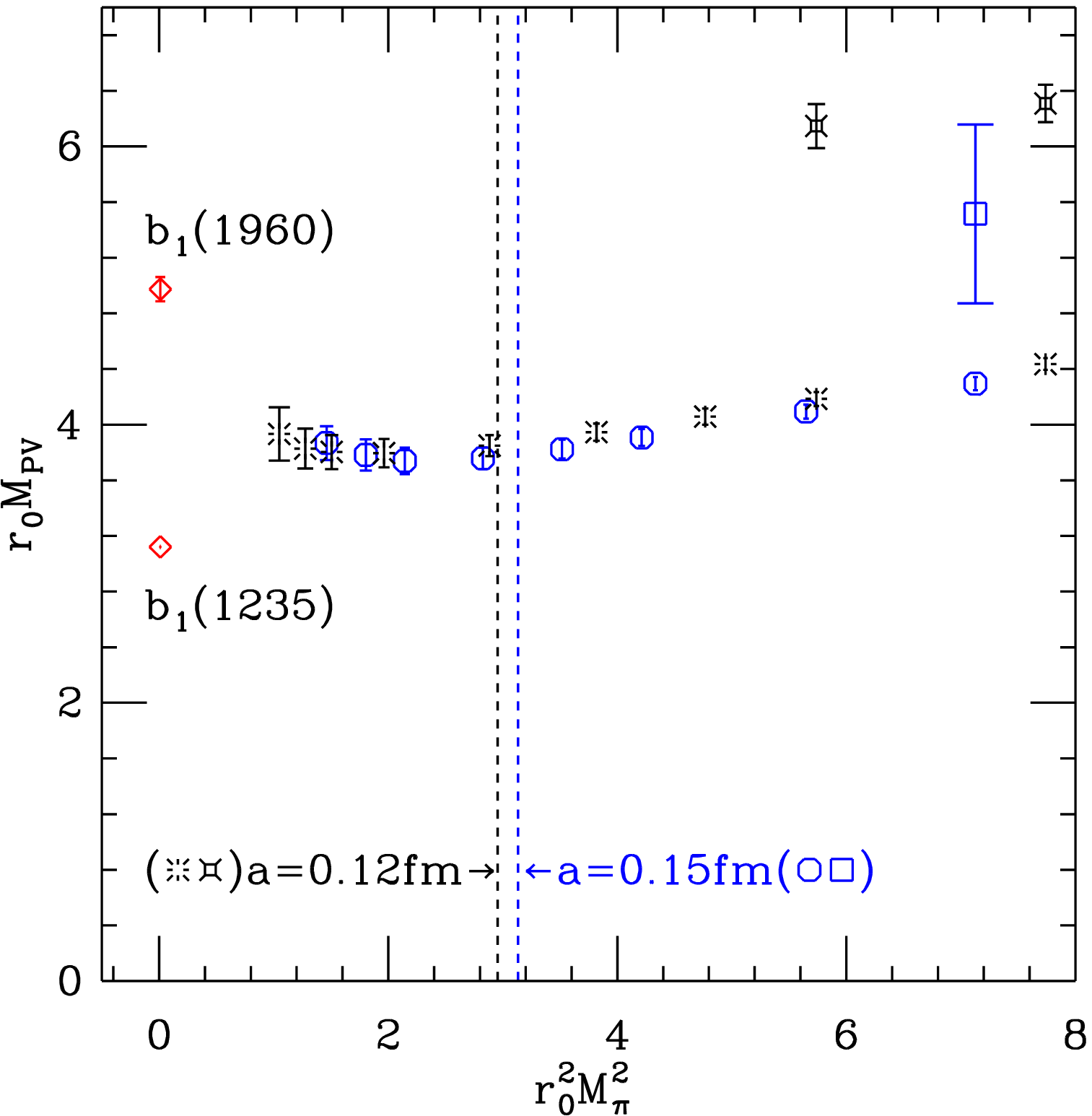}
\hspace*{5mm}
\includegraphics*[width=7.5cm]{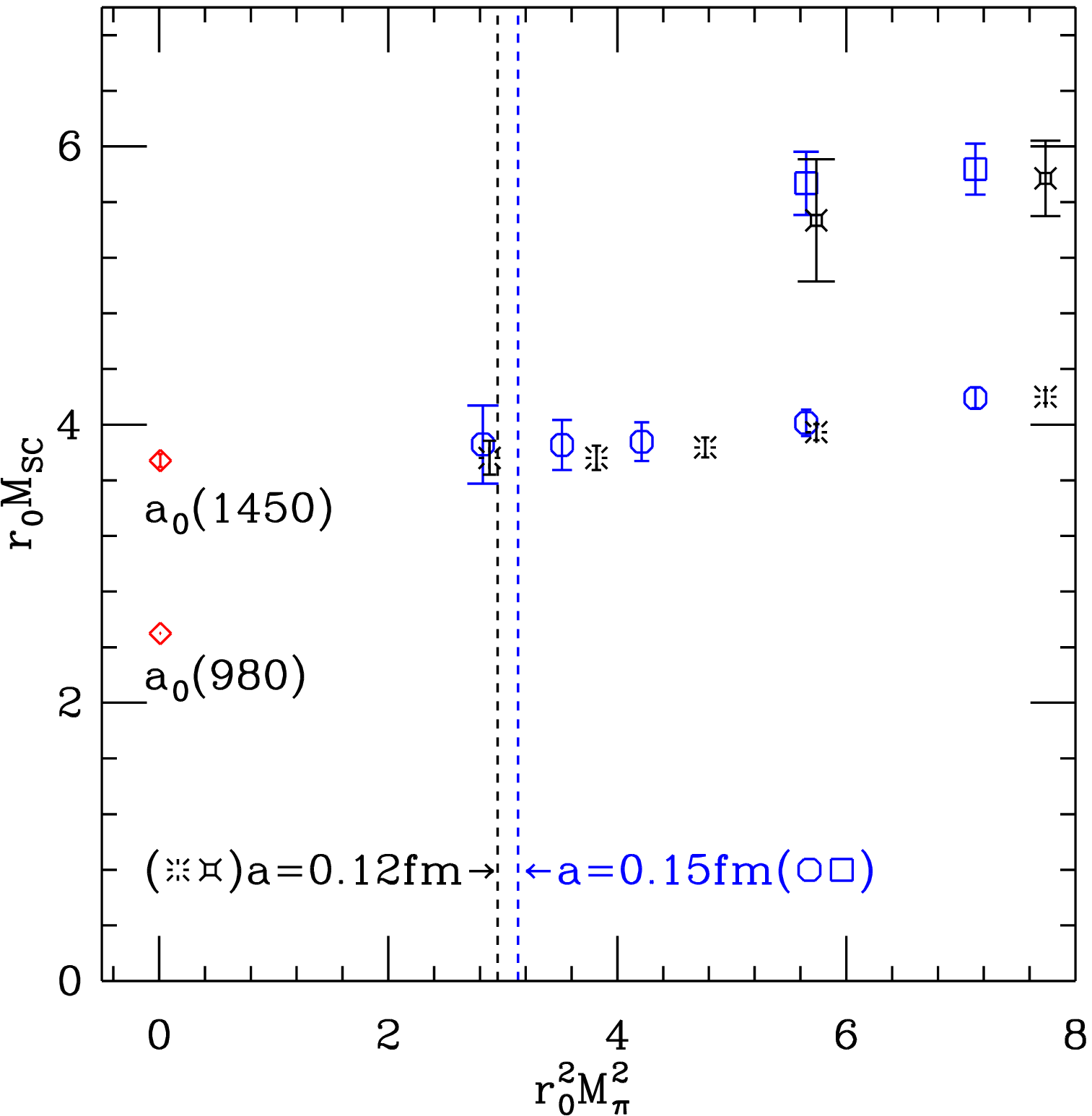}
\caption{ Ground- and excited-state meson masses vs.\ $M_\pi^2$ for both 
lattice spacings. Vector (V), pseudovector (PV), and scalar (SC) mesons 
appear. The quark masses are degenerate. 
All quantities are in units of the Sommer parameter, $r_0$. 
The diamonds represent the experimental points and the vertical lines mark 
the values of $r_0^2 M_\pi^2$ corresponding to the physical strange quark 
mass.
\label{vpts_ud_vs_ps2}}
\end{figure*}

\subsection{Quark mass dependence}

\begin{figure*}[t]
\vspace*{0mm}
\hspace*{0mm}
\includegraphics*[width=7.5cm]{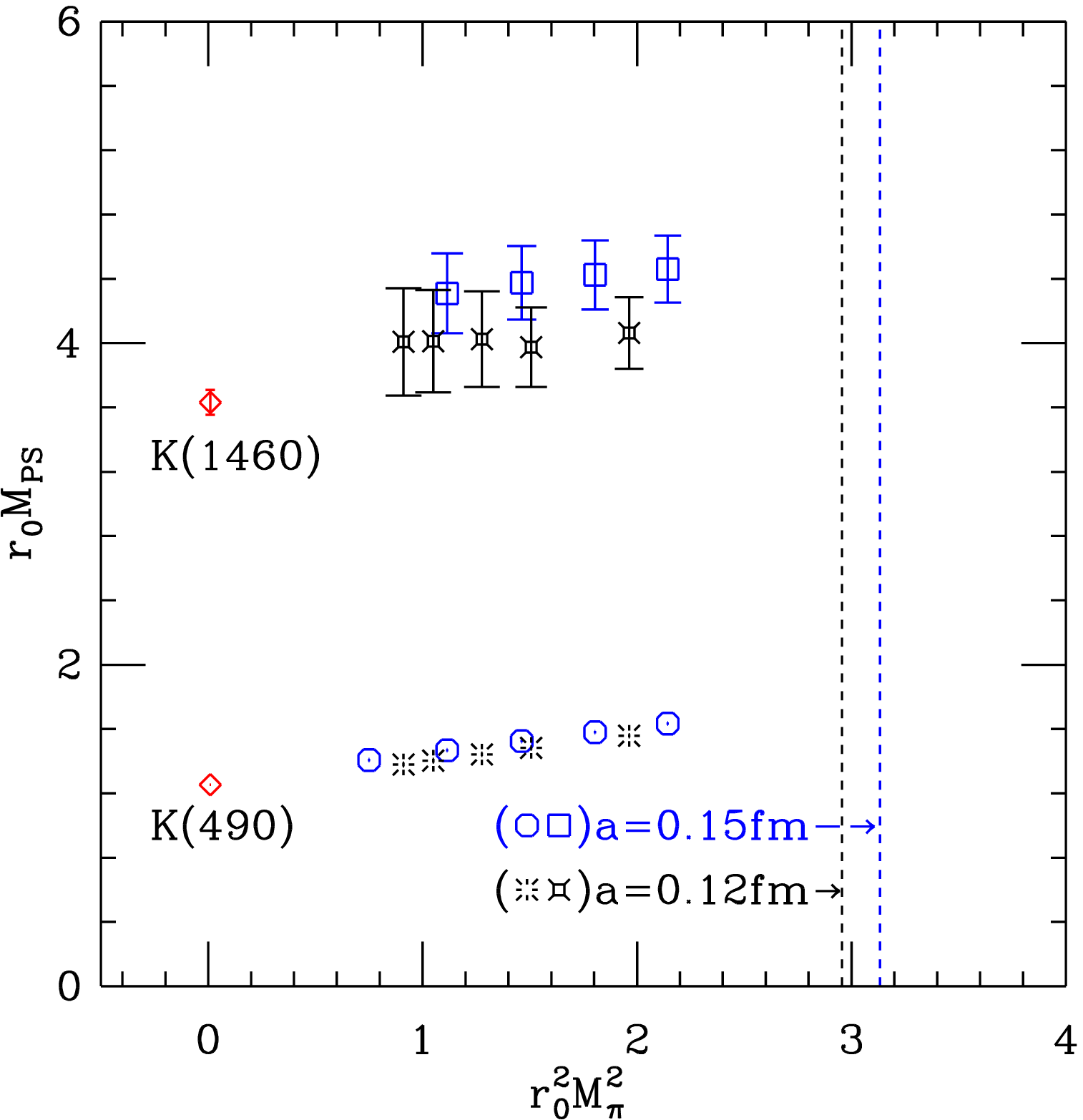}
\hspace*{5mm}
\includegraphics*[width=7.5cm]{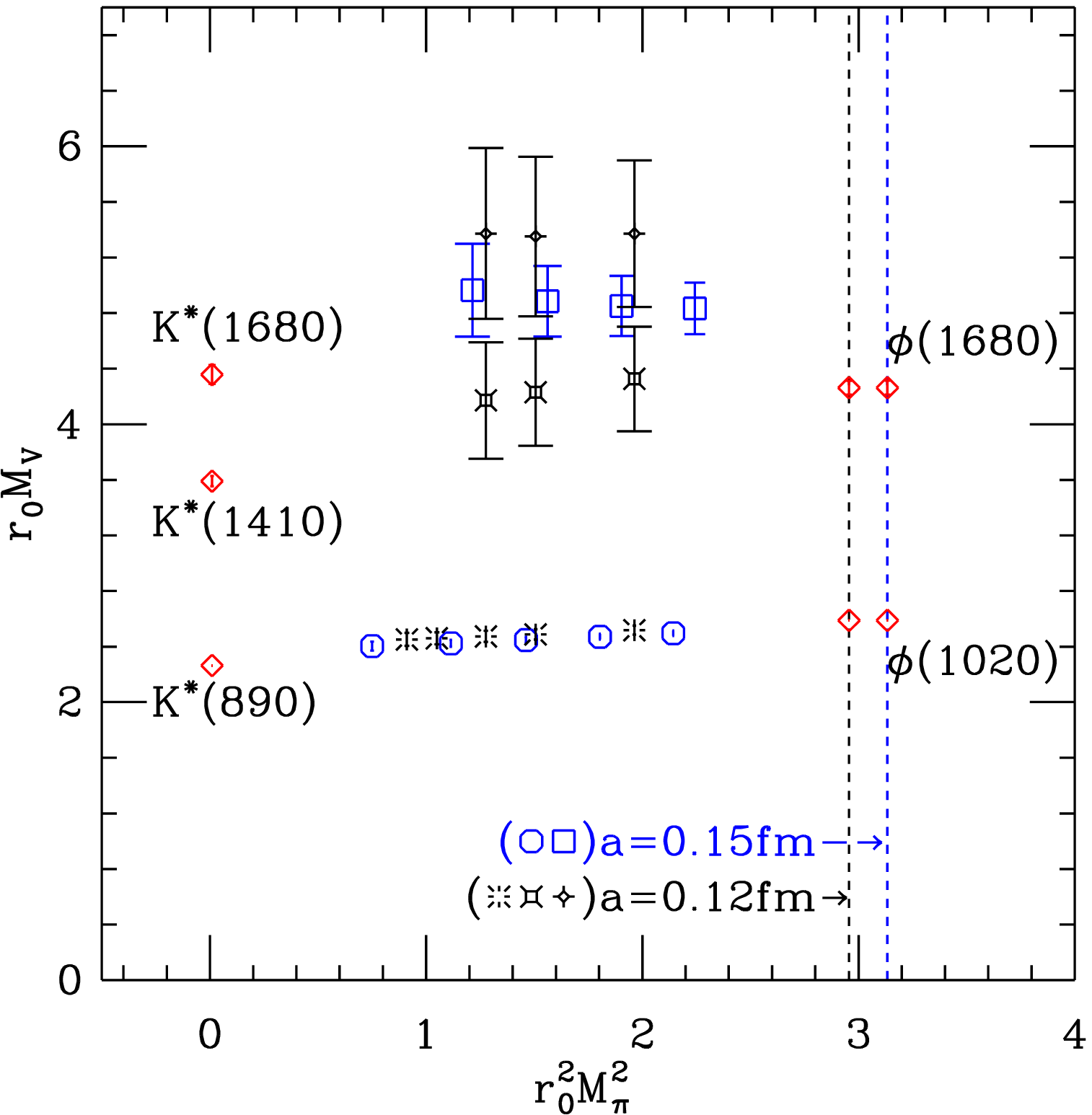}
\caption{ Ground- and excited-state meson masses vs.\ $M_\pi^2$ for both 
lattice spacings. Both pseudoscalar (left) and vector (right) mesons appear. 
One quark mass is fixed to the physical strange quark mass. 
All quantities are in units of the Sommer parameter, $r_0$. 
The diamonds represent the experimental points and the vertical lines mark 
the values of $r_0^2 M_\pi^2$ corresponding to the physical strange quark 
mass.
\label{ps_vt_us_vs_ps2}}
\end{figure*}

In Figs.\ \ref{ps_ud_vs_ps2}$-$\ref{ps_vt_us_vs_ps2}, we plot the resulting 
hadron masses versus the ground-state pseudoscalar mass-squared ($\propto m_q$), 
all in units of the Sommer parameter $r_0$. 
All plots display results from both our coarse and fine lattices. 
The vertical lines mark the values of $r_0^2 M_\pi^2$ which arise when we use 
the strange quark mass in the degenerate-quark-mass pseudoscalar; i.e., 
these mark the point where the mass of each of the light ($u,d$) quarks in 
the corresponding meson equals the strange quark mass.

Figure \ref{ps_ud_vs_ps2} shows the excited pseudoscalar masses. 
The results from both lattice sets appear consistent with the experimental 
value and, given the consistency of the two sets (within statistical errors), 
no large discretization effects are apparent.

In Fig.\ \ref{vpts_ud_vs_ps2}, we show the ground- and excited-state masses 
of the vector, pseudovector, and scalar mesons. 
For the vector mesons on the coarse lattices, only one excited state is 
extracted (another, slightly lighter effective mass plateau was seen, but 
since the eigenvectors were not stable over the same region, no fits were 
performed). 
This state either corresponds to the $\rho(1700)$ or it suffers apparently 
significant discretization effects. 
On the fine lattices, two such excited states are resolved which appear 
consistent with the physical states; the statistical errors, however, are 
larger, possibly hiding any residual systematic effects.

The pseudovector and scalar mesons present more difficulties. 
Fewer effective mass plateaus can be found for excited states; 
for the $b_1$ and scalar mesons, they are virtually non-existent. 
There are already significant problems for the ground states. 
At small quark masses, it becomes obvious that the pseudovectors suffer 
large quenching and/or finite-volume effects, causing an apparent 
enhancement of the mass.

The upward curvature seen in the ground-state vectors (at least on the coarse 
lattices) and pseudovectors may be partly explained by quenching effects. 
Quenched chiral perturbation theory for vector mesons \cite{BoChFa96} 
predicts a negative $M_\pi$ term, which could certainly cause the observed 
enhancement at low quark masses.

For the scalar meson plot similar curvature is seen, and this despite the 
fact that our method removes the dominant influences of ghosts 
\cite{BuGaGl05}. 
The entire scalar plot also appears to be shifted vertically, making one 
wonder whether the ground-state $\bar u d$ scalar meson even corresponds to 
the $a_0(980)$. Other lattice studies \cite{PrDaIz04,SuTsIs05} find that, in 
fact, it does not.

Our interpolators may be a poor choice of operators for the pseudovector 
and scalar states. 
One might achieve better results by altering the $S$-wave nature of our 
smeared quark wavefunctions. 
In a future study, we plan to apply covariant derivatives to our smeared 
sources, providing interpolators which better mimic $P$-wave orbital 
excitations and, hopefully, improving overlap with the pseudovector and 
scalar mesons.

In Fig.\ \ref{ps_vt_us_vs_ps2}, we present our results for strange-light 
pseudoscalar and vector mesons. 
The abscissa is again $r_0^2M_\pi^2$, but now this only corresponds to the 
light-quark mass and we therefore only plot values which are lighter than 
the strange-quark mass. 
The general picture here is similar to that seen for the 
degenerate-quark-mass pseudoscalar and vector mesons: 
Again, perhaps partly due to somewhat large statistical errors for the 
excited states, the results for the pseudoscalar mesons appear consistent 
with the experimental values. The same also applies to the vector mesons on 
the fine lattices, while the coarse-lattice excited vector results appear 
systematically high.

\subsection{Comparison with results on smaller lattices}

\begin{figure}[t]
\vspace*{0mm}
\hspace*{0mm}
\includegraphics*[width=7.5cm]{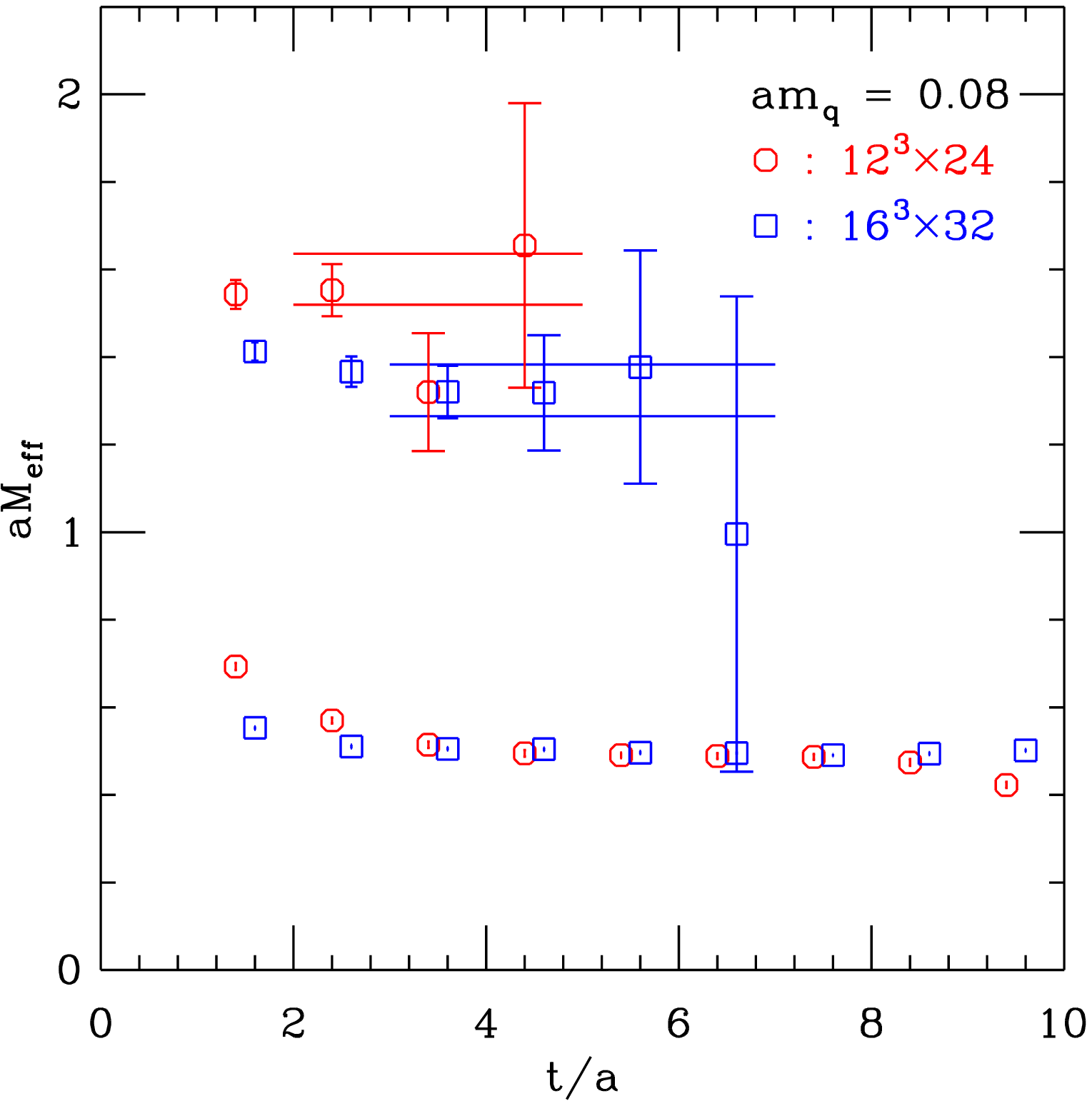}
\caption{ Ground- and excited-state effective masses for the pseudoscalars. 
Results for the $12^3 \times 24$ (see Refs.\ \cite{BuGaGl04,BuGaGl04b}) and 
$16^3 \times 32$ coarse lattices ($a \approx 0.15$ fm) are shown, along with horizontal 
lines displaying $M\pm\sigma_M^{}$ from fits to the corresponding time intervals. 
\label{effmass_12vs16}}
\end{figure}

Figure \ref{effmass_12vs16} displays two pseudoscalar effective mass plots, 
one from one of the current data sets ($16^3 \times 32$) and one from a previous 
set ($12^3 \times 24$) at the same lattice spacing 
(see Refs.\ \cite{BuGaGl04,BuGaGl04b}). 
One can see that previously chosen fit ranges (for the $12^3$ lattices) may 
have been started at too small a value of $t$, apparently enhancing the 
resulting masses when compared to our current results. 
Since the physical spatial volume has been increased from 1.8 to 2.4 fm 
(the latter perhaps still being too small for excited-state spectroscopy; 
see Ref.\ \cite{SaSa05}), 
one may be tempted to simply put this down as a finite-volume effect. 
However, since the later effective masses on the smaller lattices are 
consistent with the plateau found on the bigger ones, it is not easy to say 
whether this state is displaying effects due to the finite volume. 
The enhancement may also be due to higher state corrections, which still 
contribute since the minimum timeslice considered in the fit is too small. 
With better statistics on the $12^3$ lattices (to make up for the smaller 
number of spatial sites), we may end up finding a similar plateau to that 
seen on the $16^3$ lattices. 
Similar effects are seen for the vector mesons. 
We suspect that this poorer level of statistics for the $12^3$ lattices, 
and the subsequent choice of fit ranges, may account for most of the 
relative enhancement of our earlier results for the excited meson masses.

\subsection{Chiral extrapolations of fine lattice results}

\begin{figure}[t]
\vspace*{0mm}
\hspace*{0mm}
\includegraphics*[width=7.5cm]{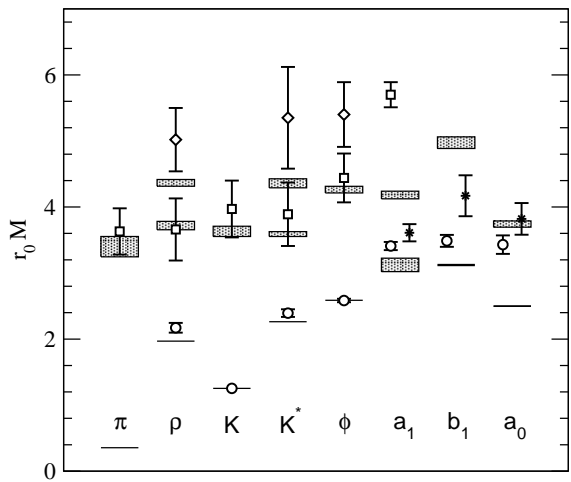}
\caption{ Chirally extrapolated (interpolated for $\phi$) results from our 
fine lattices ($20^3 \times 32$, $a=0.119$ fm). 
Most extrapolations (circles, squares, and diamonds) are linear in 
$r_0^2 M_\pi^2$. 
Those for the $a_1$, $b_1$, and $a_0$ are performed using only quark masses 
around the strange quark mass and heavier. 
The bursts on the right of the columns for these states are extrapolations 
including a $r_0 M_\pi$ term (for $a_1$ and $b_1$) or a $r_0^4 M_\pi^4$ term 
(for $a_0$), and all quark masses. 
The horizontal lines and shaded rectangles represent $M \pm \sigma_M^{}$ experimental 
values.
\label{ch_ex_mass}}
\end{figure}

Although we lack a third lattice spacing which would allow us to try to 
work in the continuum, we nevertheless attempt chiral extrapolations using 
the results from just our fine lattices (see Fig.\ \ref{ch_ex_mass}). 
For the most part, we use simple linear extrapolations (interpolations for 
$\phi$) in $r_0^2 M_\pi^2$ and, when doing so, avoid the low quark mass region 
($r_0^2 M_\pi^2 < 2.5$) for the pseudovectors since these suffer obviously 
large systematics there.

Assuming quenched chiral effects similar to those experienced by vector 
mesons \cite{BoChFa96}, we also allow for a $r_0 M_\pi$ term (obviously 
negative) in the fits for the pseudovector extrapolations and include all 
points. 
We also try a quadratic fit in $r_0^2 M_\pi^2$ for the scalars. 
These additional extrapolations are plotted as the bursts on the right of 
the column for each of these states. 
The right hand side of the plot shows the persistent difficulties of 
simulating pseudovector mesons in the quenched approximation. 
Dynamical simulations of these states appear not to suffer the same 
systematic enhancements \cite{MILC01,MILC04}. 
The extrapolations of the scalar appear more consistent with the 
$a_0(1450)$, a finding which appears consistent with unquenched results 
of simple $\bar u d$ scalars \cite{PrDaIz04,SuTsIs05}.

To the left in Fig.\ \ref{ch_ex_mass} we see that the results for the 
first-excited pseudoscalar and vector states agree with the experimental 
values (and dynamical calculations of excited pseudoscalars \cite{MILC04}), 
while the ground-state vectors (except for the $\phi$) exhibit residual 
systematic effects, a common problem in the quenched approximation.

Given the fact that our results include (possibly large) systematic errors 
(quenching, finite volume, and chiral effects) which are not presently well 
understood, we point out that the correspondence between our masses and 
the physical ones in Fig.\ \ref{ch_ex_mass} should be viewed with some 
caution. 
Nevertheless, despite these possible pitfalls, it is remarkable that the 
mass splittings appear to be of the right size.

\section{Summary}

We have presented ground- and excited-state meson masses from quenched 
lattice calculations using chirally improved quarks. 
Using a collection of linearly independent interpolating fields and the 
correlator matrix technique, we have been able to clearly separate the 
different mass eigenstates. 
Although it must be admitted that our statistical errors are still quite 
large, the masses we find for excited pseudoscalars are consistent with 
experimental values. 
The same is true for the excited vector mesons on our fine lattices. 
Results for the vectors on the coarser lattices suggest either that we have 
``missed'' the first excitation (slightly lighter effective mass plateaus 
were seen in another eigenvalue, but without corresponding plateaus for the 
eigenvector) or that this state experiences significant discretization 
effects. 
The pseudovector and scalar mesons proved more difficult to handle. 
We could obtain few results for the excited states, virtually none below 
the strange quark mass. 
At low quark masses, already the ground states have problems, displaying 
significant quenching and/or finite-volume effects. 
Ultimately, these problems will only be overcome via simulations with 
better interpolators, larger physical volumes, and dynamical fermions.

\begin{acknowledgments}
We thank Sasa Prelovsek for interesting discussions. 
The calculations were performed on the Hitachi SR8000 
at the Leibniz Rechenzentrum in Munich and we thank the LRZ staff for 
training and support. L.~Y.~G.\ is supported by ``Fonds zur F\"orderung 
der Wissenschaftlichen Forschung in \"Osterreich'', FWF, project P16823-N08. 
This work is supported by DFG and BMBF.
\end{acknowledgments}

\end{document}